\documentclass[namedreferences]{SolarPhysics}
\usepackage[optionalrh]{spr-sola-addons} % For Solar Physics 
\usepackage{graphicx}        % For eps figures, newer & more powerfull
\usepackage{color}           % For color text: \color command
\usepackage{url}             % For breaking URLs easily trough lines
\newcommand{\apj}{    {\it Astrophys. J.}}

\newcommand{\solphys}{{\it Solar Phys.}}

\begin{document}

\begin{article}

\begin{opening}

\title{Global effects of local sound-speed perturbations in the Sun: A theoretical study}

\author{S. M.~\surname{Hanasoge}$^{1}$\sep
        T. P.~\surname{Larson}$^{1}$}
\runningauthor{Hanasoge \& Larson}
\runningtitle{Global effects of local perturbations}

\institute{$^{1}$W. W. Hansen Experimental Physics Laboratory, Stanford University, Stanford, CA 94305
                     email: \url{shravan@stanford.edu} \url{tplarson@sun.stanford.edu}}

\begin{abstract}
We study the effect of localized sound-speed perturbations
on global mode frequencies by applying techniques of global helioseismology on numerical
simulations of the solar acoustic wave field. Extending the method of realization noise
subtraction (e.g. \opencite{Hanasoge07a}) to global modes and exploiting the luxury of 
full spherical coverage, we are able to achieve very highly resolved frequency differences 
that are used to study sensitivities and the signatures of the thermal asphericities. We find that
(1) global modes are almost twice as sensitive to sound-speed perturbations at the bottom of the convection zone 
as in comparison to anomalies well in the radiative interior ($r\lesssim0.55 R_\odot$), (2) the $m$-degeneracy 
is lifted ever so slightly, as seen in the $a$ coefficients, and (3) modes that propagate in the vicinity of the 
perturbations show small amplitude shifts ($\sim 0.5\%$).
\end{abstract}
\end{opening}
%-------------------------------------------------

\section{Introduction}
     \label{S-Introduction} 
Global helioseismology has proven very successful at inferring large scale
properties of the Sun (for a review, see \opencite{jcd.review}; \opencite{jcd0}). Because they
are very robust, the extension of methods of global helioseismology 
to study localized variations in the structure and dynamics of the solar interior
has been of some interest (e.g. \opencite{Swisdak}). However, the precise sensitivities of global modes 
to local perturbations are difficult to estimate through analytical means, especially in cases where 
the flows or thermal asphericities of interest possess complex spatial dependencies. To address 
questions relating to sensitivities and with the hope of perhaps discovering hitherto unknown phenomena 
associated with global modes, we introduce here for the first time a technique to study the effects of arbitrary 
perturbations on global mode parameters in the linear limit of small wave amplitudes.

Global modes attain resonant frequencies as a consequence of differentially sampling the entire 
region of propagation, making it somewhat more difficult (in comparison to local helioseismology)
to pinpoint local thermal asphericities  at depth. Exactly how difficult is one of the questions 
we have attempted to answer in this article. Jets 
in the tachocline (e.g. \opencite{jcd1}) are a subject of considerable interest 
since their existence (or lack thereof) could be very important in understanding the angular 
momentum balance of the Sun. Studying the sensitivities and signatures of waves to flows at depth 
may open up possibilities for their detection.

% Using the $a$-coefficients \cite{Schou98}, we study and place 
% bounds on the sensitivity of global modes to jets. 

Forward modeling as a means of studying wave interactions in a complex medium
like the Sun has become quite favoured (e.g. \opencite{Hanasoge06}; \opencite{Parchevsky};
\opencite{Hanasoge07a}; \opencite{Cameron}). The discovery of interesting phenomena, especially in
the realm of local helioseismology (e.g. \opencite{Hanasoge07b}; \opencite{Birch07}), adds motivation
to the pursuit of direct calculations. With the application of noise subtraction \cite{Werne, Hanasoge07a},
we can now study the signatures of a wide range of perturbations in a realistic multiple source picture. Here, we 
attempt to place bounds on the detectability of thermal asphericities at various depths in the Sun.
We introduce and discuss the method of simulation with a description of the types
of perturbations introduced in the model in Section \ref{simulations.sec}. The estimation of mode parameters 
can prove somewhat difficult due to restrictions on the temporal length of the simulation
($< 24$ hours; owing to the expensive nature of the computation). The data analysis
techniques used to characterize the modes are presented in Section \ref{peakbag.sec}.
 We then discuss the results from the analyses of the simulated
data in $\S$\ref{results.sec} and summarize this work in $\S$\ref{conclusions.sec}.

\section{Simulations and perturbations}  \label{simulations.sec}

\begin{figure} 
   \centerline{
               \includegraphics[width=0.99\textwidth,clip=]{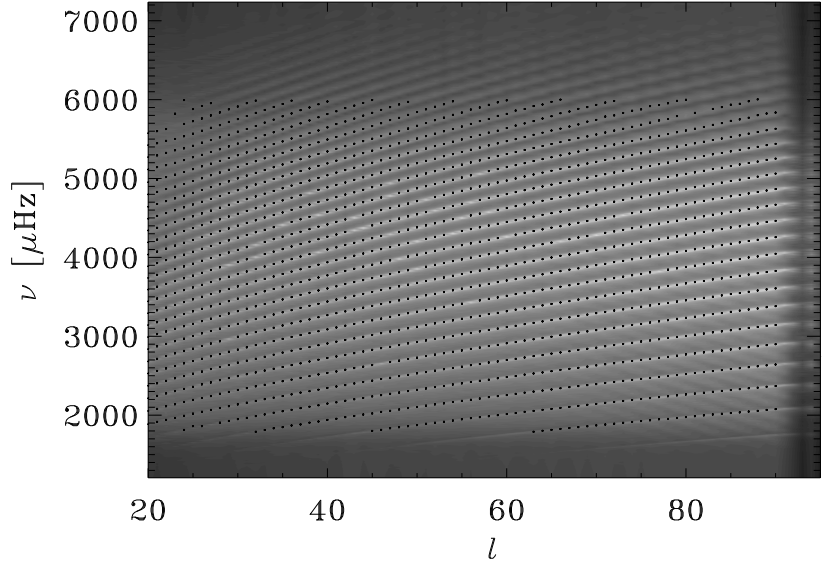}}
   \caption{Example power spectrum from a simulation with $l_{max} = 95$ and the corresponding frequency fits (symbols). This is from a
           24-hour simulation. The fitting algorithm is described in Section~\ref{peakbag.sec}. 
           Missing modes indicate that the fit for these did not converge.  We do not understand 
           why this occurs in the center of the power spectrum, but these modes can be made to converge 
           by perturbing their initial guesses. 
           Modes with $l<20$ mostly disappear from the computational domain because
           the lower boundary is placed at $r=0.24R_\odot$. We also do not excite the highest $l$ modes because 
           we wish to avoid any issues related to numerical aliasing.}\label{power.spectrum}
  \vspace{-0.2cm}
  \end{figure}

The linearized 3D Euler equations in spherical geometry are solved in the manner described in 
\inlinecite{Hanasoge06}. The computational domain is a spherical shell extending from 
$0.24 R_\odot$ to $1.002 R_\odot$, with damping sponges placed adjacent to the upper and lower
radial boundaries to allow the absorption of outgoing waves. The background stratification is
a convectively stabilized form of model S \cite{jcd.modelS, Hanasoge06}; only the highly (convectively)
unstable near-surface layers ($r > 0.98 R_\odot$) are altered while the interior is the same as model S.
 Waves are stochastically excited 
over a 200 km thick sub-photospheric spherical envelope, through the application of a dipolar 
source function in the vertical (radial) momentum equation \cite{Hanasoge06, Hanasoge07}. The forcing function 
is uniformly distributed in spherical harmonic space $(l,m)$; in frequency, a solar-like 
power variation is imposed. Any damping of the wave modes away from the boundaries is 
entirely of numerical origin. The radial velocities associated with the oscillations are extracted 
200 km above the photosphere and used as inputs to the peakbagging analyses. Data 
over the entire 360$^\circ$ extent of the sphere are utilized in the analyses, thus avoiding issues
related to mode leakage. We show an example power spectrum in Figure~\ref{power.spectrum} along with the fits.

\begin{figure}    %%%%%%%%%%%%%%%%%% FIGURE 2
                                % includes the two top panels 
   \centerline{%\hspace*{0.015\textwidth}
               \includegraphics[width=0.99\textwidth,clip=]{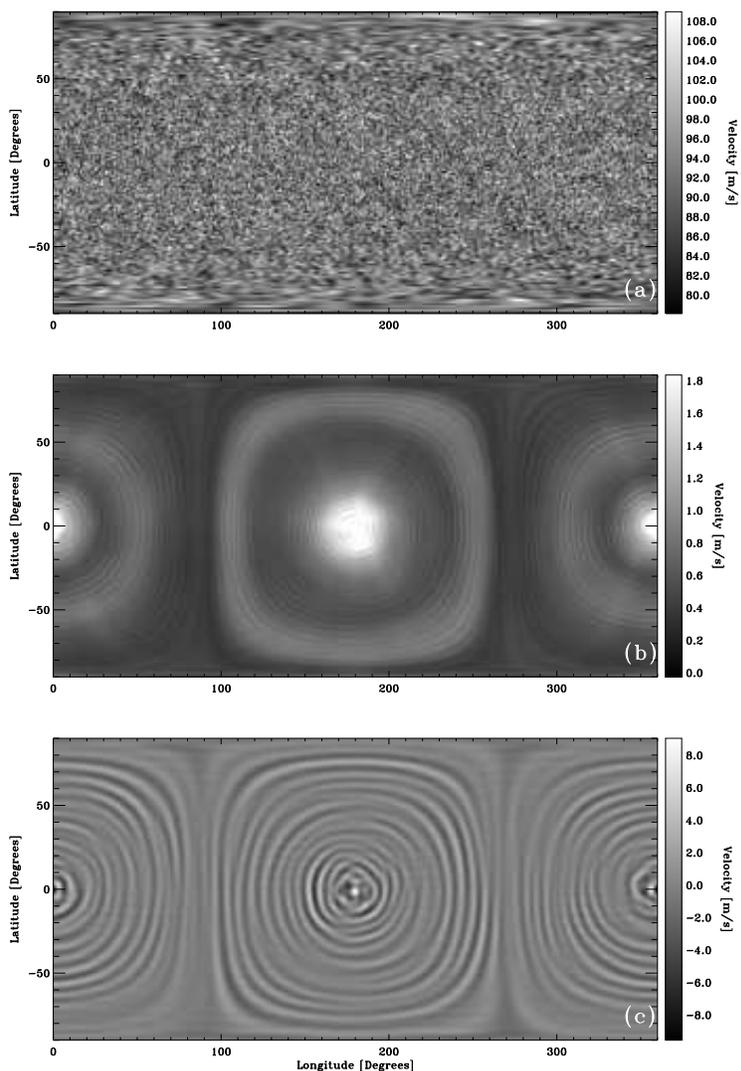}}
   \caption{Noise subtraction at work. Panel a is the time averaged RMS of the radial velocities 
of the perturbed simulation; the sound-speed perturbation (located along the equator at 
$r=0.7 R_\odot$ and 180$^\circ$ longitude) is invisible. In panel b, the time averaged 
RMS of the difference between the quiet and perturbed simulations and shown in panel c is the instantaneous 
difference. The scattering of waves and their refocusing at the antipode is clearly seen in b.}\label{noise.subtract}
  \vspace{-0.2cm}
  \end{figure}

The technique of realization noise subtraction (e.g. \opencite{Hanasoge07a}) is extensively applied in 
this work. Due to the relatively short time lengths of the simulations (the shortest time series 
yet that we have worked with is 500 minutes long!), the power
spectrum is not highly resolved and it would seem that the resulting uncertainty in the
mode parameter fits might constrain our ability to study small perturbations. To beat this limit, we perform
two simulations with identical realizations of the forcing function: a `quiet' run with no perturbations, 
and a `perturbed' run that contains the anomaly of interest. Fits to the mode parameters in these two 
datasets are then subtracted, thus removing nearly all traces of the realization and retaining 
only effects arising due to mode-perturbation interactions (see Section~\ref{peakbag.sec}). 
As an example, we show in Figure~\ref{noise.subtract} how a localized 
sound-speed perturbation placed at the bottom of the convection zone scatters waves which then proceed to refocus at 
the antipode (the principle of farside holography, \opencite{lindsey}). The presence of the sound-speed 
perturbation is not seen in panel a, whereas it is clearly seen in the noise-subtracted images 
of panels b and c. 

% Due to the relatively short time lengths of the simulation, the direct frequencies are poorly
% resolved (the shortest time series yet that we have worked with is 500 minutes long!), limiting our ability
% to look for perturbations smaller than the spread in the mode parameters. 

In these calculations, we only consider time-stationary perturbations. The sound-speed
perturbations are taken to be solely due to changes in the first adiabatic index, $\Gamma_1$; we do not 
study sound-speed variations arising from changes in the background pressure or density since altering these variables can create 
hydrostatic instabilities. Lastly, the amplitude of all perturbations are taken to be much smaller than 
the local sound speed ($\lesssim 5\%$).

% For flow perturbations, we place the requirement that the continuity 
% equation be satisfied, i.e. $\grad\cdot[\rho_0 {\bf v}] = 0$, where $\rho_0$ is the background density,
% ${\bf v}$ is the flow perturbation introduced in model, and $\grad\cdot$ is the divergence operator
% in spherical coordinates.

\section{Peakbagging analysis}
  \label{peakbag.sec}

Our first round of peakbagging is done on the $m$-averaged power spectrum for the 
quiet simulation.  For each $l$ that we attempt to fit, we search for peaks in the
negative second derivative of the power.  Unlike the power itself, which has a
background, the second derivative has the advantage of having an approximately
zero baseline.  The search is accomplished by finding the frequency at which the maximum
value of the negative second derivative occurs, estimating the mode parameters
using a frequency window of width 100 $\mu$Hz centered on this peak frequency,
zeroing the negative second derivative in this interval, and iterating. If the range
of power in the frequency window is not above a certain threshold, we check the peak
frequency found; if it is too close to a frequency found on a previous
iteration, that maximum is rejected, the same interval is again zeroed, and
iteration continues.  Note that such a simple algorithm is feasible only because
simulation data contains no leaks.  Once we have found as many peaks as possible
with this procedure, we assign a value of $n$ to each one based on a model computed
using ADIPACK \cite{jcd.adipack, Hanasoge.thesis}.

The next step is to perform an actual fit to the power spectrum in the vicinity
of each peak we identified.  For the line profile we use a Lorentzian of the
form 
\begin{equation}
P = \frac{A}{\pi} \frac{w}{(\nu - \nu _0)^2 + w^2} + B,
\end{equation}
where $A$ is the total power, $w$ is the half width at half maximum, $\nu _0$ is
the peak frequency, and $B$ is the background power.  The initial guesses for these
parameters are obtained in the first step as follows: $B$ is set to the minimum
value of the power in the frequency window around the peak, $A$ is set to the
integral under the power curve minus $B$ times the width of the window, and $w$ is
set to $1/(\pi P_{max})$ where $P_{max}$ is the maximum value of the power in the
frequency window.  The fitting interval extends halfway to the adjacent peaks, or
100 $\mu$Hz beyond the peak frequency of the modes at the edge.  The fitting itself
is done using the IDL routine {\texttt{curvefit}}.

Once we have fit these mode parameters for the $m$-averaged spectrum, we use them as 
the initial guesses for fitting the individual $m$ spectra.  Then for each $l$ and $n$
we can fit a set of $a$-coefficients to the frequencies as functions of $m/l$. 
Although for the quiet sun we would expect for all the $a$-coefficients to be zero,
this calculation is still necessary in order to perform the noise subtraction.

We also use the mode parameters from the $m$-averaged spectrum of the quiet simulation 
as initial guesses for fitting the (unshifted) $m$-averaged spectrum of the perturbed
simulation.  Although the perturbations may lift the degeneracy in $m$, we expect
the splitting to be very small, so that the peaks in the $m$-averaged spectrum can
still be well represented by a Lorentzian. We also use those same initial guesses
for fitting the individual $m$ spectra of the perturbed simulation, and 
recalculate the $a$-coefficients.

%Although in principle the error in the difference between the mode parameters for the
%two different simulations is due entirely to the finite numerical precision of the
%simulation,
An empirical estimate of the error in frequency differences for the sound-speed
perturbation at $r=0.7 R_\odot$ (see Section~\ref{sound.speed.anomalies}) 
is computed in the following manner. We look at the difference
in mode parameters only for those modes that do not penetrate to the depth of the
perturbation (all modes with $\nu/(l+1/2) < 60$).  We then make a histogram of these
differences with a bin size of 0.001 $\mu$Hz and fit a Gaussian to the resulting
distribution. With this method we find a standard deviation of 0.000474 $\mu$Hz
or 0.47 nHz. This result is confirmed by also computing the standard deviation of 95\% of the closest points
to the mean.
 
\section{Results and discussion}
   \label{results.sec}
\subsection{Sound-speed anomalies}\label{sound.speed.anomalies}
\begin{figure}    
   \centerline{%\hspace*{0.015\textwidth}
               \includegraphics[width=0.99\textwidth,clip=]{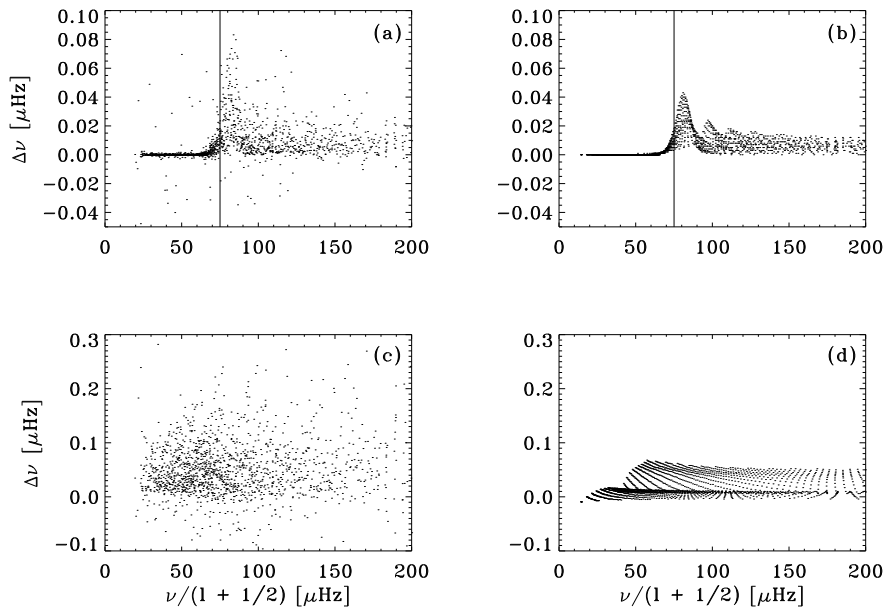}}
   \caption{Frequency shifts ($\Delta\nu$) plotted against phase speed ($\nu/[l+1/2]$) for the sound-speed perturbations of Section~\ref{sound.speed.anomalies}. Solid lines indicate the phase speed of waves that have $r=0.7 R_\odot$ as inner turning points. Panel a shows the shifts due to a localized hot spot (sound-speed increase) at the bottom of the convection zone. It appears that modes whose inner turning points are just below the bottom of the convection zone are the ones maximally sensitive to the perturbation. Panel b shows the frequency shifts obtained from ADIPACK for the spherically symmetric component of this perturbation. In panel c, all modes feel the presence of the relatively large near-surface hot spot and in d, the shifts predicted by ADIPACK for the spherically symmetric analog for this near-surface anomaly are shown.  Noise subtraction does not remove the realization noise associated with the scattering process itself; therefore the spread in the frequency shifts of the simulated data is greater than ADIPACK ones. }\label{sound.speeds}
  \vspace{-0.2cm}
  \end{figure}

We place three perturbations of horizontal size $8^\circ \times 8^\circ$ (in longitude and latitude) with a full width at half maximum in radius of 2$\% R_\odot$ (13.9 Mm) at depths of $r = 0.55, 0.7, 1.0 R_\odot$, each with an amplitude $\alpha$ of $+5$\% of the local sound speed. Because of the fixed angular size, the perturbations grow progressively smaller in  physical size with depth; our intention was to keep the perturbation as localized and non-spherically symmetric as possible. Despite the fact that the perturbation is highly sub-wavelength (the wavelength at $r=0.7 R_\odot$ is $76$ Mm or $11\% R_\odot$), we notice that for these (relatively) small amplitude anomalies, the global mode frequency shifts are predominantly a function of the spherically symmetric component of the spatial structure of the perturbation. In other words, what matters most is the contribution from the $l=0$ coefficient in the spherical harmonic expansion of the horizontal spatial structure of the perturbation. We verify this by computing the frequency shifts associated with a spherically symmetric area-averaged version of the localized perturbation (with an amplitude of $0.05 A^{local}/(4\pi)$, where $A^{local}$ is the solid angle subtended by the localized perturbation, 0.05 referring to the 5\% increase in sound speed). We were careful to ensure that the radial dependence of the magnitude of the perturbation was unchanged. The frequency shifts associated with the spherically symmetric perturbations were calculated independently through simulation and the oscillation frequency package, ADIPACK \cite{jcd.adipack} and seen to match accurately, as shown in Figure~\ref{sound.speeds}.

\begin{figure}    
   \centerline{%\hspace*{0.015\textwidth}
               \includegraphics[width=0.99\textwidth,clip=]{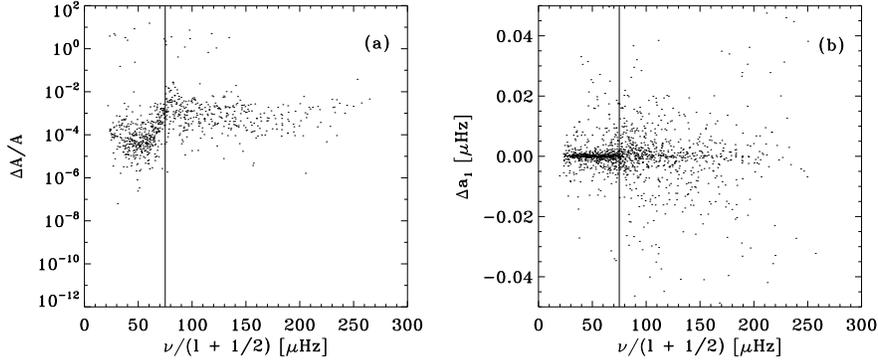}}
   \caption{Changes in mode amplitudes (panel a) and the $a_1$ coefficients (panel b) due to the sound-speed perturbation located at $r = 0.7 R_\odot$ shown as functions of the phase speed of the waves. 
In panel a, it seen that only modes which have turning points close to and below the location of the 
perturbation show changes in the amplitude (on the order of 0.5\% or so). Due to spatially localized nature of the sound-speed 
perturbation, the $m$ degeneracy is lifted, creating the slight bump in the $a_1$ coefficient. Although not
shown here, we observe that several $a$ coefficients, even and odd, show the presence of the perturbation.}\label{mode.parameters.fig}
  \vspace{-0.2cm}
  \end{figure}

Because of the non-spherically symmetric nature of the perturbation, we expect to see shifts in the $a$-coefficients. 
Similarly, it is likely that there will be slight deviations in the amplitudes of modes that propagate in regions close to and below the locations of the perturbation. We display these effects for the case with the perturbation located at $r = 0.7 R_\odot$ in Figure~\ref{mode.parameters.fig}. 

\subsection{Scattering extent}
We introduce a non-dimensional measure, $\kappa$, to characterize the degree of scattering exhibited by the anomaly:
\begin{equation}
\kappa = \frac{4\pi}{\alpha A} \sqrt{\frac{1}{N}\sum_{n,l,m} \left(\frac{\delta\nu}{\nu}\right) ^2}\label{kappa.eq}, %\right],
\end{equation}
where $\alpha = \delta c/c$, the amplitude of the sound-speed perturbation expressed in fractions of the local sound speed, $A$ the angular area of the perturbation, and $N$ the number of modes in the summation term. Essentially, this parameter tells us how strongly perturbations couple with the wave field, with larger $\kappa$ implying a greater degree of scatter and vice versa. Because it is independent of perturbation size or magnitude, $\kappa$ can be extended to study flow perturbations as well. This measure is meaningful only in the regime where the frequency shifts are presumably linear functions of the perturbation magnitude. Also, it is expected that $\kappa$ will retain a strong dependence on the radial location of the perturbation since different parts of
the spectrum see different regions of the Sun. For example, placing an anomaly at the surface will likely affect the entire spectrum of global modes, as seen in Figure~\ref{sound.speeds}c. Results for $\kappa$ shown in Table~\ref{T-scatter} contain no surprises; for a given size and magnitude of the perturbation, the effect on the global frequencies increases strongly with its location in radius. The signature of a perturbation at the bottom of the convection zone on the global modes is twice as strong as an anomaly in the radiative interior ($r=0.55 R_\odot$). The surface perturbation is a little more difficult to compare with the others because contrary to the two deeper perturbations, it is locally far larger than the wavelengths of the modes. The result however is in line with expectation; the near-surface scatterer is far more potent than the other two anomalies.

% and $v/c$ in the case of the jet
% Essentially, with the introduction of $\kappa$, we can estimate and directly compare
% $\kappa^{\rm J}$ (jet) and $\kappa^{\rm SS}$ (sound-speed anomaly) to determine which one couples better with the 
%modes.

%  \begin{figure}    %%%%%%%%%%%%%%%%%% FIGURE 2
                                % includes the two top panels 
%   \centerline{%\hspace*{0.015\textwidth}
%               \includegraphics[width=0.99\textwidth,clip=]{jplot.eps}}
%   \caption{Frequency shifts due to a jet at the bottom of the convection. Shown are shifts in 
%            the $m$-averaged specrum as a function of frequency (panel a) and phase speed (panel b), 
%            changes in $a_1$ (panel c), and the fit to the frequency as a function of $m$ (panel d). We do not
%            have an explanation for the dramatic frequency variation of panel a. It is seen that waves that
%            have inner turning points above the location of the jet are affected in the opposite manner as those
%            which plumb greater depths. A faint bump is observed in the $a_1$ coefficient shifts in panel c. The
%            amplitude of the jet velocity is 4.56 km/s, corresponding to frequency of rotation of $\sim 1~\mu$Hz.}\label{jet.shifts}
%  \vspace{-0.1cm}
%  \end{figure}

\begin{table}
\caption{The scattering extents $\kappa$, of various perturbations. The Root
Mean Square (RMS) variation in frequencies is shown as well.
}
\label{T-scatter}
\begin{tabular}{ccclc}     % define the column alignment
                           % l: left, c: center, r: right
  \hline                   % horizontal line
Depth & RMS & $\kappa$ \\
     ($r/R_\odot$)  & $\delta\nu/\nu$ &  \\
  \hline
0.55 & $0.20\times 10^{-5}$  & 0.023 \\
0.70 & $0.44\times 10^{-5}$  & 0.049  \\
1.00 & $2.07\times 10^{-5}$  & 0.231 \\
%Jet & 0.71 & $64.2\times 10^{-5}$  & 1.621 \\
  \hline
\end{tabular}
\end{table}

\section{Conclusion}
   \label{conclusions.sec}
We have introduced a method to systematically study the effects of various local perturbations on global mode frequencies. Techniques of mode finding and parameter fitting are applied to artificial data obtained from simulations of wave propagation in a solar-like stratified spherical shell. We are able to beat the issue of poor frequency resolution by extending the method of realization noise subtraction \cite{Hanasoge07a} to global mode analysis. These methods can prove very useful in the study of shifts due to perturbations of magnitudes beyond the scope of first order perturbation theory; moreover, extending this approach to investigate systematic frequency shifts in other stars may prove exciting. We are currently studying the impact of complex flows like convection and localized jets on the global frequencies. Preliminary results seem to indicate that flows are stronger scatterers (larger $\kappa$) than sound-speed perturbations although more work needs to be done to confirm and characterize these effects. 

\acknowledgements
  S. M. Hanasoge and T. P. Larson were funded by grants HMI NAS5-02139 and MDI NNG05GH14G. We
would like to thank Jesper Schou, Tom Duvall, Jr., and Phil Scherrer for
useful discussions and suggestions. The simulations were performed on the
Columbia supercomputer at NASA Ames.

%Moreover, one can study the effects of
%arbitrary perturbations on global frequencies to develop a better understanding of the
%interactions that

\end{article}

\end{document}